\documentclass[twocolumn,showpacs,preprintnumbers,amsmath,amssymb]{revtex4}

\usepackage{amsmath,amssymb,graphicx}
\newcommand{\bea}{\begin{eqnarray}}
\newcommand{\eea}{\end{eqnarray}}

\begin{document}

\vspace*{-10ex} \hspace*{\fill} TCC-006-09
\title{Infrared divergence of pure Einstein gravity contributions to
       cosmological density power spectrum}
\author{Hyerim Noh${}^{1}$, Donghui Jeong${}^{2,3}$ and Jai-chan Hwang${}^{4}$}
\address{${}^{1}$Korea Astronomy and Space Science Institute,
                 Daejon, Korea \\
         ${}^{2}$Department of Astronomy, University of Texas at
                 Austin, University Station, C1400, Austin, TX 78712,
                 USA \\
         ${}^{3}$Texas Cosmology Center, University of Texas at
                 Austin, University Station, C1400, Austin, TX 78712,
                 USA \\
         ${}^{4}$Department of Astronomy and Atmospheric Sciences,
                 Kyungpook National University, Taegu, Korea \\
         E-mails: ${}^{1}$hr@kasi.re.kr,
                  ${}^{2}$djeong@astro.as.utexas.edu,
                  ${}^{4}$jchan@knu.ac.kr}

\begin{abstract}

We probe the pure Einstein's gravity contributions to the
second-order density power spectrum. In the small-scale, we {\it
discover} that the Einstein's gravity contribution is negligibly
small. This guarantees that Newton's gravity is sufficient to handle
the baryon acoustic oscillation scale. In the large scale, however,
we {\it discover} that the Einstein's gravity contribution to the
second-order power spectrum dominates the linear-order power
spectrum. Thus, pure Einstein gravity contribution appearing in the
third-order perturbation leads to an infrared divergence in the
power spectrum.

\end{abstract}

\noindent 

\maketitle

\vskip .0cm
%
%
The large-scale cosmological density power spectrum can be regarded
as one of the main pillars of modern cosmology where theories meet
with observations. Recent discovery of the baryon acoustic
oscillation (BAO) near $100h^{-1}\mathrm{Mpc}$ scale \cite{BAO} has
spurred renewed interests in cosmology community on the importance
of detailed theoretical studies of the physics behind the BAO
including the initial spectrum and the nonlinear process; $h$ is a
Hubble constant in the unit of $100km/sec/Mpc$. The BAO provides an
important distance scale of sound horizon size at the baryon
decoupling epoch, which is completely set by the anisotropic angular
power spectrum of cosmic microwave background radiation (CMB), and
hence, can be used as a \textit{standard ruler}. This standard ruler
encoded in the power spectrum provides the measurement of the
angular diameter distance and the Hubble expansion rate by which we
can constrain the properties of dark energy. Recent studies show
that nonlinear (next-to-leading order) processes in the power
spectrum is important in theoretical analysis of the BAO phenomenon
especially in the context of planned precise observations of the
high redshift galaxy surveys \cite{Jeong-Komatsu-2008}.

Up until now, however, all studies of the cosmological power spectra
are treated in the context of Newton's gravity. The linear-order
cosmological perturbation was first handled in Einstein's gravity by
Lifshitz in 1946 \cite{Lifshitz-1946}. Later a Newtonian study was
made by Bonnor in 1957 \cite{Bonnor-1957}, and it was shown that
Newtonian density perturbation equation coincides exactly with the
relativistic result in the zero-pressure limit. We have recently
shown in \cite{second-order} that the density perturbation equation
in Einstein's gravity coincides exactly with the previously known
result in Newton's gravity \cite{Peebles-1980} even to the
second-order: we call it a relativistic/Newtonian correspondence to
the second-order perturbation. This is a nontrivial result which is
available in the temporal comoving gauge with suitable
identifications of relativistic metric and matter variables. In the
zero-pressure case the Newtonian hydrodynamics is known to be closed
to the second order in nonlinearity \cite{Peebles-1980}. Thus, any
nonvanishing third-order terms in Einstein's gravity can be
naturally identified as pure general relativistic contributions.
Again with the same comoving gauge and a suitable choice of
variables recently we have derived the density perturbation equation
with the third-order corrections \cite{third-order}; all variables
in our gauge condition are naturally gauge-invariant.

The pure Einstein's gravity third-order terms contribute to the
next-to-leading-order (second-order) density power spectrum. In this
{\it Letter}, we present contribution of the pure Einstein's gravity
corrections to density power spectrum. Our results reveal
cosmologically important discoveries in both the small scale and the
large-scale. Particularly surprising is the unexpected surfacing of
an infrared divergence due to the general relativistic nonlinear
effect in the large scale which potentially demands a new revised
understanding of the current standard paradigm of the large-scale
structure formation mainly based on the linear perturbation theory.

\vskip .0cm
%
%
We expand the density fluctuation as $\delta \equiv \delta_1 +
\delta_2 + \delta_3 + \cdot \cdot \cdot$ where $\delta ({\bf x}, t)
\equiv \delta \varrho ({\bf x}, t)/\varrho(t)$ is the relative
density fluctuation. The density power spectrum becomes \bea
   & & P \equiv \langle |\delta (k,t) |^2 \rangle
       = \langle |\delta_1|^2 \rangle
       + \langle |\delta_2|^2 \rangle
       + 2 \langle Re(\delta_1^* \delta_3) \rangle
       + \cdot \cdot \cdot
   \nonumber \\
   & & \quad
       \equiv P_{11} + P_{22} + P_{13} + \cdot \cdot \cdot ,
\eea where $k$ is a wave number in Fourier space and $\langle
\rangle$ indicates phase space averaging; $\langle Re(\delta_1^*
\delta_2) \rangle$ vanishes assuming the random phase
\cite{Vishniac-1983}; a superscript $*$ indicates a complex
conjugation. The density power spectrum is introduced as
   $
       \langle \delta ({\bf k}) \delta ({\bf k}^\prime) \rangle
       \equiv (2 \pi )^3 \delta_D ({\bf k} + {\bf k}^\prime) P(k)
   $
with $k = |{\bf k}| = |{\bf k}^\prime|$ and $\delta_D$ a Dirac
delta-function, \cite{quasilinear}. The $P_{11}$ is the linear-order
power spectrum, and $P_{22} + P_{13}$ is the second-order
(next-to-leading order) power spectrum. We notice that the
third-order pure general relativistic correction terms contribute to
$P_{13}$; thus we decompose it as $P_{13} = P_{13,\rm Newton} +
P_{13,\rm Einstein}$. The density and velocity power spectra valid
to the next-to-leading-order are presented in
\cite{NH-power-spectrum-2008}. An integration of the density power
spectrum in Eq.\ (19) of \cite{NH-power-spectrum-2008} gives
\begin{widetext} \bea
   & & \langle | \delta (k, t) |^2 \rangle
       = \langle | \delta_1 (k,t) |^2 \rangle
       + {1 \over 98} {k^3 \over (2\pi)^2}
       \int_0^\infty dr
       \left| \delta_1 (kr, t) \right|^2
       \int_{-1}^1 dx
       \left| \delta_1
       \left( k \sqrt{1 + r^2 - 2 r x}, t \right) \right|^2
       {\left( 3 r + 7 x - 10 r x^2 \right)^2 \over
       \left( 1 + r^2 - 2 r x \right)^2}
   \nonumber \\
   & & \quad
       + {1 \over 252} {k^3 \over (2 \pi)^2}
       \left| \delta_1 (k, t) \right|^2
       \int_0^\infty dr
       \left| \delta_1 (kr, t) \right|^2
       \bigg[
       - 42 r^4
       + 100 r^2
       - 158
       + {12 \over r^2}
       + {3 \over r^3} \left( r^2 - 1 \right)^3
       \left( 7 r^2 + 2 \right)
       \ln{\left| {1 + r \over 1 - r} \right|}
       \bigg]
   \nonumber \\
   & & \quad
       +{10 \over 21}
       \left( {\ell \over \ell_H} \right)^2
       {k^3 \over (2 \pi)^2}
       \left| \delta_1 (k, t) \right|^2
       \int_0^\infty dr \left| \delta_1 (kr, t) \right|^2
       \bigg[
       - {41 \over 6} r^2
       - 21 - {45 \over 4} {1 \over r^2}
   \nonumber \\
   & & \qquad \qquad
       + {9 \over 8}
       {1 \over \left( 1 + r \right)^2 \left( 1 - r \right)^2}
       \left( 5 r^6 - 13 r^4 + 9 r^2
       + 1 - {2 \over r^2} \right)
       +{3 \over 16}
       \left( 43 r^2 + 46 r - {53 \over r}
       - {36 \over r^3} \right)
       \ln{ \left| {1 - r \over 1 + r} \right|}
       \bigg]
   \nonumber \\
   & & \quad
       \equiv P_{11} + P_{22} + P_{13,\rm Newton} + P_{13,\rm
       Einstein},
   \label{PS}
\eea \end{widetext} where $r \equiv k^\prime/ k$ and $x\equiv ({\bf
k} \cdot {\bf k}^\prime)/(kk^\prime )$; ${\ell / \ell_{H}} \equiv
{\dot a / (kc)}$ is the ratio between a scale $\ell \equiv a/k$ and
the horizon scale $\ell_H \equiv c/H$ with $H \equiv \dot a/a$, and
$a(t)$ the cosmic scale factor. The second-order power spectrum in
Eq.\ (\ref{PS}) is derived in Einstein's gravity; as we have shown
that results in Einstein's gravity {\it coincide} exactly with the
Newton's one to the second-order perturbations, the power spectrum
in Eq.\ (\ref{PS}) without $P_{13, \rm Einstein}$ is exactly valid
in Newton's gravity. The $P_{22}$ and $P_{13, \rm Newton}$ were
presented in \cite{Suto-Sasaki-1991} in the Newtonian context. The
pure Einstein's gravity contribution $P_{13, \rm Einstein}$ is the
new result in this work; in Eq.\ (\ref{PS}) we assumes flat matter
dominated background. This pure Einstein's gravity contribution to
the one-loop corrected power spectrum is hitherto unknown in the
literature based on Newton's gravity \cite{quasilinear}. We note
that the pure Einstein's gravity contribution is multiplied by a $(
{\ell / \ell_H} )^2$ factor, thus suppressed far inside the horizon.
The relativistic/Newtonian correspondence to the second order is
valid in the zero-pressure situation without rotation, no
gravitational waves, and spatially flat background, but valid in the
presence of the cosmological constant; for the general cases, see
\cite{Multi}.

\vskip .0cm
%
%

%
%
\begin{figure*}
\begin{center}
\rotatebox{90}{%
\includegraphics[width=9.1cm]{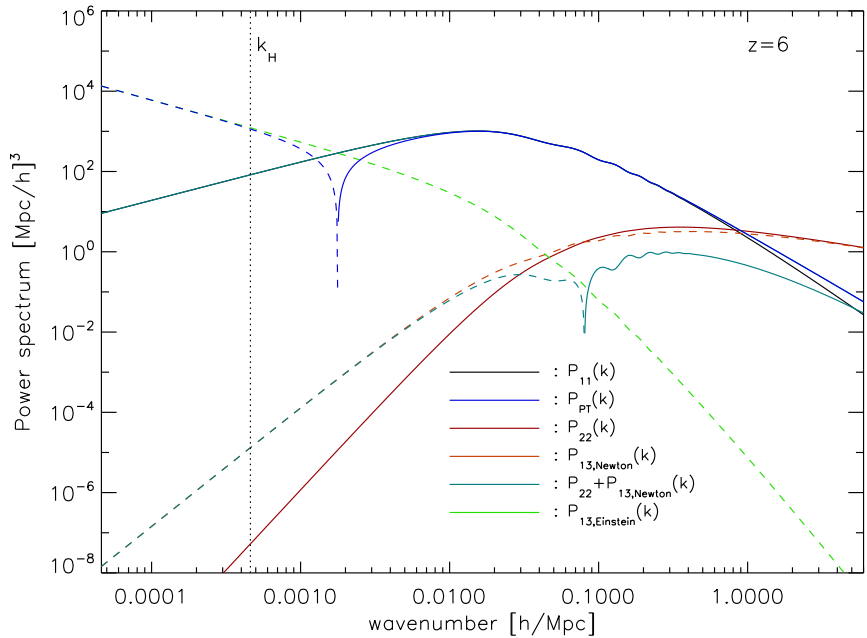}
}%
\caption{
         Second-order power spectrum and the contribution from each
         component of Eq.\ (2) at $z=6$. Note that we take the absolute value
         for negative terms, and show with dashed lines. Vertical dotted line
         shows the wavenumber corresponding to the current comoving horizon
         ($k_H$). We use $k_{min}=10^{-2}k_H$ and $k_{max}=10^3$ to evaluate
         the integration in Eq.\ (2).
            [\textit{color online}]
         }%
\label{fig1}
\end{center}
\end{figure*}

In Figure \ref{fig1} we present integration of equation (\ref{PS})
by using a realistic linear power spectrum from a concordence
cosmology of $\Lambda$CDM Universe. We calculate the linear power
spectrum using the {\sf CAMB} \cite{CAMB} code with the maximum
likelihood cosmological parameters given in Table 1 of
\cite{komatsu/etal:2008} (``WMAP+BAO+SN"). Since equation (\ref{PS})
is valid for the spatially flat, matter dominated universe, we
calculate the second (next-to-leading) order power spectrum at
higher redshift ($z=6$) when our Universe is well approximated by
flat, matter dominated universe. When calculating the integration,
we set the integration lower bound as $r_\mathrm{min}=10^{-2}k_H/k$,
where $k_H=1/\ell_H$ is the wave number corresponding to the horizon
scale, and upper bound as $r_\mathrm{max}=10^3/k$. Note that, unlike
the Newtonian correction terms, $P_{22}$ and $P_{13,\rm Newton}$,
$P_{13,\rm Einstein}$ is sensitive to $r_{max}$, and logarithmically
diverges. It is a problem that the amplitude of the Einstein term
contains a logarithmically divergent integral; a logarithmic
divergence
is also encountered in \cite{Kolb-etal-2005} in considering the
backreaction of inhomogeneity on the cosmological expansion
\cite{referee-comments}.

In the small-scale it was found in \cite{Vishniac-1983} that the
leading order of $P_{22}$ {\it cancels} exactly with the leading
order of $P_{13,\rm Newton}$. Our result in Figure \ref{fig1} shows
that despite such a cancelation of the leading-order Newtonian
contributions, the next leading order Newtonian contribution is
still bigger than the pure general relativistic contribution
$P_{13,\rm Einstein}$. We find that $P_{13,\rm Einstein}$ is roughly
$1\%$ of $P_{22}+P_{13,\rm Newton}$ at $k=0.1~h/\mathrm{Mpc}$, and
it becomes smaller and smaller as $k$ increases. This may be a good
news to the cosmology community based on Newton's gravity because
our result guarantees to use Newton's gravity in handling the weakly
nonlinear processes far inside the horizon including the BAO scale.
In the large scale, however, Figure \ref{fig1} reveals a completely
unexpected surprising result: the pure general relativistic
contribution to the second-order power spectrum dominates over the
linear-order relativistic/Newtonian power spectrum. We emphasize
that previous nonlinear perturbation studies based on Newton's
gravity have been mainly concerned with the small-scale effect where
indeed the fluctuations grow from linear to nonlinear phases. We
will discuss implications of this new discovery in the following.

\vskip .0cm
%
%
Figure \ref{fig1} apparently shows that $P_{13,\rm Einstein}$ is
bigger than even the linear-order density power spectrum $P_{11}$ in
the large scale; $P_{13,\rm Einstein}$ is negative in general. It is
not necessarily a problem that the Einstein term dominates at the
smallest wavenumber $k$; this also happens for instance when a
fluctuation distribution with no (very little) large scale power at
higher order grows a $k^4$ tail at small $k$
\cite{Zeldovich-1965,referee-comments}. It is, however, a problem
that the dominant contribution is negative, which is a sign that
perturbation theory is at least incomplete; the full power spectrum
is of course positive definite \cite{referee-comments}. At its face
value our result implies that as the second-order effect is larger
than the linear one the nonlinear effect of Einstein's gravity leads
to breakdown of the perturbation theory in the large scale but far
inside the horizon scale. The situation is particularly awkward
because in the equation level the third-order pure Einstein's
gravity correction terms are smaller than the second-order
relativistic/Newtonian terms by a factor $\varphi_v \sim \delta
\Phi/c^2$, see Eqs.\ (1)-(3) in \cite{NH-power-spectrum-2008};
$\delta \Phi$ is the Newtonian gravitational potential which is
generally quite small, $\delta \Phi /c^2 \sim 10^{-7} - 10^{-5}$.
Thus, we naturally anticipate the perturbatively derived third-order
solutions from the relativistic/Newtonian second-order equations
[these are smaller than the second-order terms by a factor $\delta
\sim (\ell_H/\ell)^2 \delta \Phi$] are bigger than the pure
Einstein's gravity contributions inside the horizon, see Eq.\
(\ref{PS}). That is, in the large-scale limit ($k \rightarrow 0$ and
$r \rightarrow \infty$) a naive examination of Eq.\ (\ref{PS}) shows
that both $P_{13,\rm Newton}$ and $P_{13,\rm Einstein}$ have
$k$-dependence proportional to
       $k^{-2}
       \left| \delta_1 (k, t) \right|^2
       \int_0^\infty dk^\prime (\dots)$.
However, an asymptotic expansion of $P_{13,\rm Newton}$ shows that
both the leading-order term ($-42 r^4$) and even the
next-to-leading-order term ($100 r^2$) {\it cancel} away, whereas no
such cancelation occurs in $P_{13,\rm Einstein}$. This explains the
diverging $k$-dependence of $P_{13,\rm Einstein}$ in Figure
\ref{fig1} compared with the behavior of $P_{13,\rm Newton}$. The
divergence of $P_{13,\rm Einstein}$ is not a real problem as the
density fluctuation is described by ${\cal P} \equiv (k^3/2 \pi^2)
P$; as we have $P_{13,\rm Einstein} \propto k^{-1}$, ${\cal
P}_{13,\rm Einstein}$ is convergent in that limit
\cite{referee-comments}.

A similar naive examination of Eq.\ (\ref{PS}) in the small scale
leads us to expect that $P_{13,\rm Einstein}$ could be comparable to
the Newtonian second-order contributions. That is, in the
small-scale limit ($k \rightarrow \infty$ and $r \rightarrow 0$) a
naive examination of Eq.\ (\ref{PS}) shows that $P_{13,\rm Newton}$
has a $k$-dependence proportional to
       $k^4
       \left| \delta_1 (k, t) \right|^2
       \int_0^\infty dk^\prime (\dots)$,
whereas $P_{13,\rm Einstein}$ has a $k$-dependence proportional to
       $k^2
       \left| \delta_1 (k, t) \right|^2
       \int_0^\infty dk^\prime (\dots)$.
However, as the leading-order terms in $P_{13,\rm Newton}$ {\it
cancel} exactly with the one in $P_{22}$ \cite{Vishniac-1983}, we
have $P_{22} + P_{13,\rm Newton}$ has the same $k$-dependence as the
leading-order term ($-45/(4r^2)$) in $P_{13,\rm Einstein}$. Thus, we
naturally anticipate that $P_{13,\rm Einstein}$ could be comparable
with the relativistic/Newtonian contribution to the second-order
power spectrum in the small scale. However, an asymptotic expansion
of $P_{13,\rm Einstein}$ shows that both the leading-order term
($-45/(4r^2)$) and even the next-to-leading-order term ($-21$) {\it
cancel} away, whereas no such cancelation occurs in $P_{13,\rm
Newton}$. This explains why $P_{13,\rm Einstein}$ is far smaller
than the Newtonian contributions in the small scale.

As a matter of fact, we can hardly be able to accept the result that
the Einstein's gravity leads to breakdown of the perturbation
expansion in the scale where perturbation amplitude is known to be
near linear. Being confronted by this unexpected situation we 
have gone through all the algebra several times without
finding a mistake in the analytic calculations leading to equation
(\ref{PS}); for convenience in the close examination of the algebra
by the readers we made the whole calculation in a pdf file available
in the web \cite{third-order-pdf}; the numerical code is also
uploaded in the web \cite{code}.
Besides any potential error made in our part, we point out
that the infrared divergence could be due to an improper choice of
our gauge condition. Our results are based on the temporal comoving
gauge, setting $\tilde T^0_\alpha \equiv 0$. The
relativistic/Newtonian correspondence to the second order is
available only under this gauge. We have continued to use the same
gauge condition and identification of perturbation variables
\cite{third-order}. Whether we could find a nondivergent power
spectrum in other gauge condition is an interesting possibility to
be investigated in the future work.

\vskip .0cm
%
%
Recently, Losic and Unruh presented an intriguing possibility of
divergent behaviors of second-order perturbations in the quantum
generation stage during inflation which might have a close relation
to our result \cite{Losic-Unruh-2008}. They argued that ``a certain
nonlocal measure of second-order metric and matter perturbations
generically dominates in its amplitude compared to that of the
linear-order perturbations during slow-roll inflation''. They
further argued that ``during slow roll, second-order fluctuations
grew large for a class of inflationary models'' and ``nonlinear, and
probably nonperturbative, gravitational effects dominate near
slow-roll spacetime, and therefore the linear perturbation theory
likely fails in those situations''. The linear power spectrum used
in Figure \ref{fig1} is based on a near Harrison-Zel'dovich spectrum
\cite{Harrison-1970} which naturally arises from quantum
fluctuations during the slow-roll inflation era
\cite{Mukhanov-Chibisov-1981}; this is often regarded as the major
triumph of the inflation scenario in the early universe and helped
to make inflation scenario a firm theoretical feature in the early
universe despite its energy scale far beyond experimentally
reachable range. Our result in this {\it Letter} is analogous to and
consistent with the argument made by Losic and Unruh. Based on the
near Harrison-Zel'dovich spectrum, we find a possible divergence in
the large scale due to pure Einstein's gravity effect on the
second-order power spectrum. One way out of this conundrum is to
have a correct initial power spectrum generated in the inflation era
including the role of the second or higher order perturbation theory
during slow-roll inflation era.

Divergent results found in the nonlinear perturbations in the large
scale of density power spectrum here and in the seed generation
stage of early inflation era found by Losic and Unruh, if confirmed
to be correct, seriously challenge the currently accepted standard
paradigm of theoretical cosmology concerning the cosmological
structure formation theory and physics in the early universe. A
systematic investigation of the second-order perturbation theory
during the quantum generation stage based on inflation is needed to
resolve the issue raised in \cite{Losic-Unruh-2008}. Our result also
suggests systematic investigations to be made concerning the role of
third-order perturbations on the CMB temperature anisotropy power
spectrum in the large scale.

\vskip .0cm
%
%
\noindent{\bf Acknowledgments:} We thank Professor Robert
Brandenberger for useful discussion.
J.H.\ was supported by the Korea Research Foundation (KRF) Grant
funded by the Korean Government (MOEHRD, Basic Research Promotion
Fund) (No.\ KRF-2007-313-C00322) (KRF-2008-341-C00022), and by Grant
No.\ R17-2008-001-01001-0 from the Korea Science and Engineering
Foundation (KOSEF). D.J.\ was supported by Wendell Gordon Endowed
Graduate Fellowship of the University of Texas at Austin. H.N.\ was
supported by grants No.\ C00022 from KRF and No.\ 2009-0078118 from
KOSEF funded by the Korean Government (MEST).

%
%


\begin{thebibliography}{99}
\bibitem{BAO}
         D. J. Eisenstein {\it et al.} \emph{Ap. J.} {\bf 633}, 560 (2005).
\bibitem{Jeong-Komatsu-2008}
         D. Jeong and E. Komatsu [astro-ph] arXiv:0805.2632.
\bibitem{Lifshitz-1946}
         E. M. Lifshitz \emph{J. Phys.} (USSR) \textbf{10}, 116 (1946).
\bibitem{Bonnor-1957}
         W. B. Bonnor \emph{Monthly Not. R. Astron. Soc.} \textbf{117}, 104 (1957).
\bibitem{second-order}
         H. Noh and J. Hwang \emph{Phys. Rev. D} {\bf 69}, 104011 (2004).
\bibitem{Peebles-1980}
         P. J. E. Peebles \emph{The large-scale structure of the universe}
         (Princeton Univ. Press, Princeton) (1980).
\bibitem{third-order}
         J. Hwang and H. Noh \emph{Phys. Rev. D} {\bf 72}, 004012
         (2005);
         \emph{Monthly Not. R. Astron. Soc.} {\bf 367}, 1515 (2006).
\bibitem{Vishniac-1983}
         E. T. Vishniac \emph{Monthly Not. R. Astron. Soc.} {\bf 203}, 345 (1983).
\bibitem{quasilinear}
         J. N. Fry \emph{Astrophys. J.} {\bf 421}, 21 (1994);
         F. Bernardeau, S. Colombi, E. Gaztanaga and R. Scoccimarro
         \emph{Phys. Rep.} {\bf 367}, 1 (2002).
\bibitem{NH-power-spectrum-2008}
         H. Noh and J. Hwang \emph{Phys. Rev. D} {\bf 77}, 123533 (2008).
\bibitem{Suto-Sasaki-1991}
         Y. Suto and M. Sasaki \emph{Phys. Rev. Lett.} {\bf 66}, 264 (1991).
\bibitem{Multi}
         J. Hwang and H. Noh \emph{Phys. Rev. D} {\bf 76}, 103527 (2007).
\bibitem{CAMB}
         A. Lewis, A. Challinor and A. Lasenby \emph{Astrophys. J.} {\bf 538}, 473 (2000).
\bibitem{komatsu/etal:2008}
         E. Komatsu, {\it et al.} ArXiv e-print, {\bf 803} (2008).
\bibitem{Kolb-etal-2005}
         E. W. Kolb, S. Matarrese, A. Notari and A. Riotto \emph{Phys. Rev. D}
         {\bf 71}, 023524, (2005).
\bibitem{referee-comments}
         This comment is suggested by an anonymous referee.
\bibitem{Zeldovich-1965}
         Ya. B. Zel'dovich, \emph{Adv. Astron. Ap.} {\bf 3}, 241 (1965);
         P. J. E. Peebles, \emph{Astron. Astrophys.} {\bf 32}, 391 (1974);
         S. F. Shandarin and A. L. Melot \emph{Astrophys. J.} {\bf 364}, 396 (1990).
\bibitem{third-order-pdf}
         H. Noh and J. Hwang
         http://bh.knu.ac.kr/$\sim$jchan/third-order-note.pdf
\bibitem{code}
         J. Jeong https://webspace.utexas.edu/dj955/www/codes.htm
\bibitem{Losic-Unruh-2008}
         B. Losic and W. G. Unruh \emph{Phys. Rev. Lett.} {\bf 101}, 111101 (2008).
\bibitem{Harrison-1970}
         E. R. Harrison \emph{Phys. Rev. D} {\bf 1}, 2726 (1970);
         Ya. B. Zel'dovich \emph{Monthly Not. R. Astr. Soc.} {\bf 160}, 1 (1972).
\bibitem{Mukhanov-Chibisov-1981}
         V. Mukhanov and G. Chibisov \emph{JETP Lett.} {\bf 33},
         532-536 (1981);
         S. Hawking \emph{Phys. Lett. B} {\bf 115}, 295 (1982);
         A. Starobinsky \emph{Phys. Lett. B} {\bf 117}, 175 (1982);
         A. Guth and S. Y. Pi \emph{Phys. Rev. Lett.} {\bf 49}, 1110 (1982);
         J. Bardeen, P. Steinhardt and M. Turner \emph{Phys. Rev. D} {\bf 28},
         679 (1983).
\end{thebibliography}
\end{document}